\documentclass[aps,longbibliography,superscriptaddress,eqsecnum,twocolumn,nofootinbib]{revtex4-2}
\usepackage{graphicx}
\usepackage{stmaryrd}
\usepackage{amsmath, amsfonts,amssymb}
\usepackage{mathrsfs}
\usepackage{braket}
\usepackage{color}
\usepackage{xspace}
\usepackage{amscd}
\usepackage{comment}
\usepackage{bm}
\usepackage{bbm}
\definecolor{lightblue}{rgb}{0.13, 0.26, 0.99}

\usepackage[
colorlinks=true,
urlcolor=blue,
citecolor=blue,
linkcolor=blue,
hyperfootnotes=false]{hyperref}

\newcommand{\im}{\operatorname{Im}}
\newcommand{\re}{\operatorname{Re}}

\allowdisplaybreaks

\usepackage{color}
\usepackage[normalem]{ulem}

\begin{document}

\title{Spin pumping into quantum spin chains}
\author{Shunsuke C. Furuya}
\affiliation{Department of Liberal Arts, Saitama Medical University, Moroyama, Saitama 350-0495, Japan}
\affiliation{Institute for Solid State Physics, The University of Tokyo, Kashiwa, 277-8581, Japan}
\author{Mamoru Matsuo}
\affiliation{Kavli Institute for Theoretical Sciences, University of Chinese Academy of Sciences, Beijing, 100190, China}
\affiliation{CAS Center for Excellence in Topological Quantum Computation, University of Chinese Academy of Sciences, Beijing, 100190, China}
\affiliation{Advanced Science Research Center, Japan Atomic Energy Agency, Tokai, 319-1195, Japan}
\affiliation{RIKEN Center for Emergent Matter Science (CEMS), Wako, Saitama, 351-0198, Japan}
\author{Takeo Kato}
\affiliation{Institute for Solid State Physics, The University of Tokyo, Kashiwa, 277-8581, Japan}

\date{\today}
\begin{abstract}
We theoretically investigate spin pumping into a quantum easy-plane ferromagnetic spin chain system. This quantum spin chain is effectively described by the Tomonaga-Luttinger (TL) liquid despite the ferromagnetic exchange interaction because of the easy-plane magnetic anisotropy. 
This TL liquid state has an extremely strong interaction that is hardly realized in other quantum antiferromagnetic chain systems or weakly interacting electron systems.
We show how the strongly interacting TL liquid affects the ferromagnetic resonance that occurs in the ferromagnetic insulator.
In particular, we discuss the dependence of the Gilbert damping on the temperature and the junction length.
The Gilbert damping allows us to extract information about the above-mentioned strong interaction within the quantum ferromagnetic spin chain.
We also point out that a well-known compound CsCuCl$_3$ will be suitable for the realization of our setup.
\end{abstract}

\maketitle

\section{Introduction}\label{sec:intro}

Spin pumping~\cite{Tserkovnyak2002,simanek2003,tserkovnyak2005rmp,Hellman2017}, the spin current generation through magnetization dynamics driven by ferromagnetic resonance (FMR), is actively investigated in spintronics as a versatile way to efficiently inject spins into adjacent materials~\cite{Zutic2004,Tsymbal2019}. 
The spin pumping can also be regarded as a method of detecting spin excitations, since the injected spin current reflects the information about spin excitations~\cite{Han2020,Yang2018,Qiu2016,Yamamoto2021,Ominato2020a,Ominato2020b,Yama2021,Inoue2017,Silaev2020,Silaev2020b,simensen2021,fyhn2021,Ominato2022a,Ominato2022b,Funato2022,sun2023uSC,funaki2023,sun2023altermagnet,Yama2023,Fukuzawa2023}.

Application of the spin pumping technique to low-dimensional materials is one of the challenges that are appealing, since they exhibit characteristic spin excitations related to, e.g., topological phases, Dirac electron systems, and quantum Hall systems~\cite{Rojas-Sanchez2013_2d, Shiomi2014_2d, Rojas-Sanchez2016_2d, Gong2017_2d,Huang2017_2d,Fei2018_2d,Zhang2020_2d,Burch2018_vdW_review}.
It is also remarkable that the spin pumping experiment has interface sensitivity, which has advantages for detection of spin excitation in pure 2D systems such as atomic layer compounds, compared with nuclear magnetic resonance (NMR) and neutron scattering experiment.
Among the characteristic quantum phases in low-dimensional spin systems, the Tomonaga-Luttinger (TL) liquid is one of the most typical phenomena~\cite{Voit1995,vonDelft1998,Giamarchi2003}.
In one-dimensional materials such as carbon nanotubes and various quantum spin chain materials, low-energy spin excitation is described by the TL liquid.
The TL liquid shows a critical scaling law in various observable quantities, interestingly, with non-universal power-law behaviors.
Unlike many other critical phenomena, the power law of scaling in the TL liquid reflects the information of the strength of the interaction among constituent elements of the TL liquid.
Such a characteristic critical behavior is stable against changes of parameters, \textit{e.g.},  temperatures.
Hence, the one-dimensional systems described by the TL liquid, which has an infinitely long correlation length along the chain, will eventually lead to the large signal strength in a wide range of temperatures.

Fukuzawa \textit{et al.} recently discussed spin pumping into the carbon nanotube~\cite{Fukuzawa2023}, where the linewidth of the FMR turned out to give us information about the electron-electron interaction in the carbon nanotube.
The FMR linewidth is directly related to the Gilbert damping of the ferromagnet.
Therefore, the interaction strength of the adjacent carbon nanotube governs the scaling law of the Gilbert damping of the ferromagnet through the interface interaction between them.

An increase or decrease in the interaction strength will change the power-law behavior of the Gilbert damping and eventually alter the characteristics of the spin pumping.
Thus, it will be worth pursuing a one-dimensional material that exhibits an interesting spin pumping phenomenon.
Quantum spin chain materials are a promising candidate for a platform providing exotic spin excitations even in the field of one-dimensional quantum systems for a reason we show soon later.
The magnetic anisotropy of the easy plane determines the interaction strength of the TL liquid in quantum spin chains~\cite{Giamarchi2003}.

\begin{figure}[t!]
\centering\includegraphics[bb = 0 0 753 354, width=85mm]{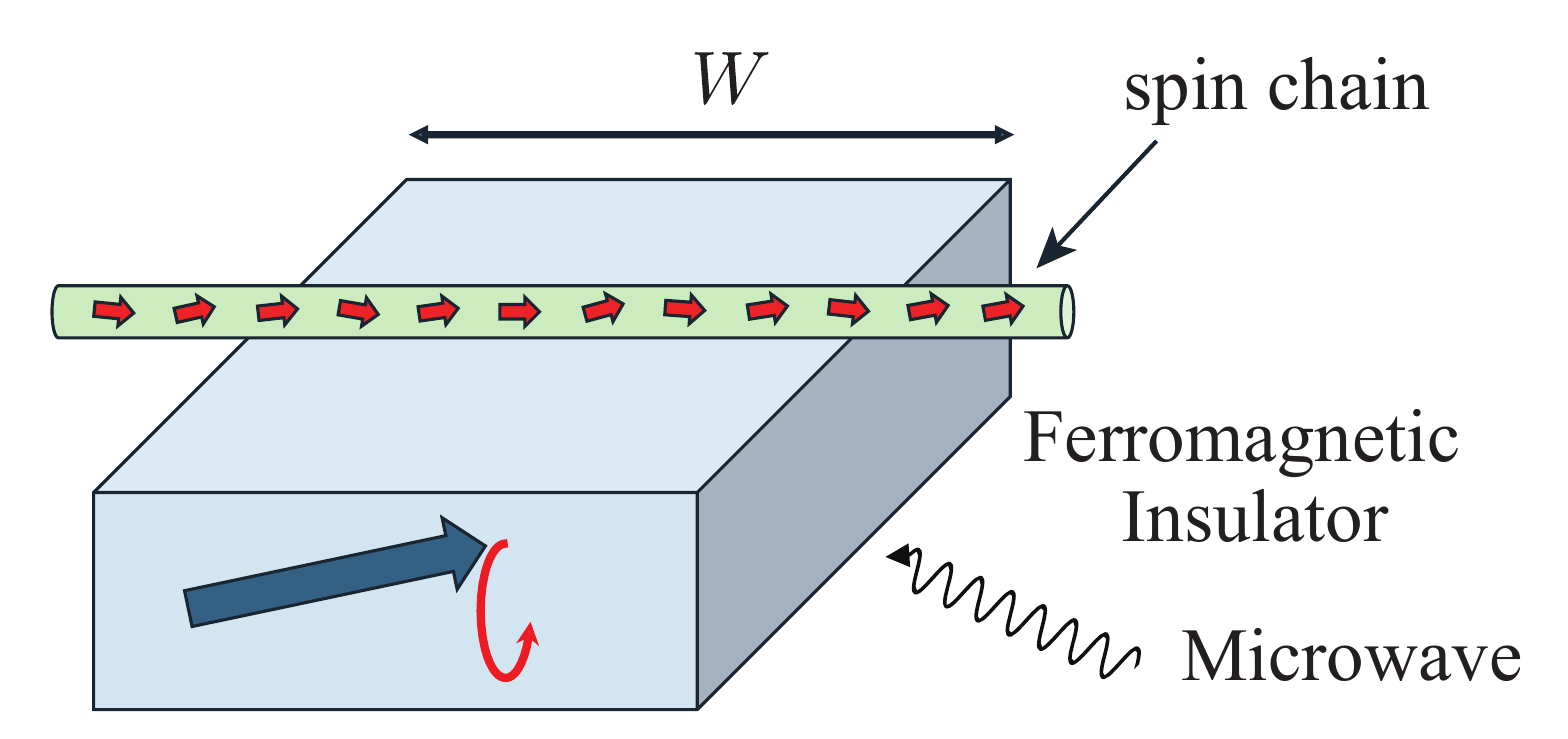}
    \caption{Schematic figure of the junction system composed of a three-dimensional ferromagnetic insulator and a quasi-one-dimensional ferromagneic insulator. For simplicity, one spin chain in the quasi-one-dimensional magnet is depicted.
    We consider the ferromagnetic resonance (FMR) experiment for the three-dimensional ferromagnet by applying microwave in the presence of the uniform magnetic field.}
    \label{fig:setup}
\end{figure}

In this paper, we discuss the spin pumping from the (three-dimensional) ferromagnetic insulator to the one-dimensional quantum spin chain material (Fig.~\ref{fig:setup}).
As a promising candidate, we consider an interesting quantum spin chain material of CsCuCl$_3$~\cite{Rioux1969_CsCuCl3,Adachi1980_CuCsCl3,Hyodo1981_CsCuCl3,Tanaka1992_CsCuCl3_ESR,Yamamoto2021_CsCuCl3,Nihongi2022_CsCuCl3}.
CsCuCl$_3$ is a quantum \emph{ferromagnetic} chain with an easy-plane magnetic anisotropy.
We show that the TL liquid of such a quantum spin chain leads to a large interaction strength and eventually to the characteristic temperature dependence of the Gilbert damping.

We organize this paper as follows.
Section~\ref{sec:model} describes our theoretical model of spin pumping and the theoretical framework of the spin pumping.
We note that our model has a simple feasible Hamiltonian.
We review the electron spin resonance (ESR) of the spin chain and the FMR of the three-dimensional ferromagnet, where we also introduce the main object of this paper, the Gilbert damping.
Section~\ref{sec:FM} gives the main part of our theoretical analyses without going into deep technical details.
Section~\ref{sec:material} discusses an important topic of candidate material to realize our model.
We summarize the paper in Sec.~\ref{sec:summary}.

\section{Model}\label{sec:model}

Our theory widely applies to quantum spin chains with the easy-plane anisotropy, though we keep in mind the specific material, CsCuCl$_3$.
In this section, we formulate our theory as general as possible and refer to the material when necessary.

\subsection{Hamiltonian}\label{sec:hamiltonian}

Our model Hamiltonian consists of three parts:
\begin{align}
    \mathcal H_{\rm tot}
    &= \mathcal H_{\rm 3d} +\mathcal H_{\rm 1d} + \mathcal H_{\rm int}.
    \label{H_tot_def}
\end{align}
$\mathcal H_{\rm 3d}$, $\mathcal H_{\rm 1d}$, and $\mathcal H_{\rm int}$ are the bulk Hamiltonian of the ferromagnetic insulator, the bulk Hamiltonian of the quantum spin chain material, and an interfacial interaction between them, respectively.
We suppose that the bulk ferromagnetic insulator is described by a quantum ferromagnetic Heisenberg model on a three-dimensional lattice:
\begin{align}
    \mathcal H_{\rm 3d} = J_{\rm F} \sum_{\braket{i,j}} \bm S_i \cdot \bm S_j-\hbar \gamma_g h_{\rm dc} \sum_i S_i^z,
    \label{H_FI_def}
\end{align}
with $J_{\rm F}<0$.
The operator $\bm S_i$ is a spin-$S_0$ operator, denoting the local spin in the ferromagnetic insulator.
$\gamma_g$ and $h_{\rm dc}\ge 0$ are the gyromagnetic ratio and the uniform static magnetic field, respectively.
The lattice can be anything as long as the spontaneous ferromagnetic order along the $S^z$ direction emerges at low temperatures.

The spin-chain Hamiltonian is an XXZ-type,
\begin{align}
    \mathcal H_{\rm 1d}/N_c &= J \sum_i (s_i^xs_{i+1}^x + s_i^y s_{i+1}^y + \Delta_z s_i^z s_{i+1}^z) 
    \notag \\
    &\qquad -\hbar \gamma'_g h_{\rm dc} \sum_i s_i^z,
    \label{H_1d_def}
\end{align}
where $\bm s_i$ is the spin-$1/2$ operator of the spin-chain material and $J$ represents the intrachain exchange coupling.
$\Delta_z$ is the uniaxial magnetic anisotropy.
We assume that the magnetic field is weak enough so that $\hbar \gamma'_g h_{\rm dc}\ll |J|$
and $\Delta_z$ takes a value in the range,
\begin{align}
    -1< \Delta_z < 1.
    \label{Delta_z_range}
\end{align}
Then, the ground state of the Hamiltonian \eqref{H_1d_def} is the TL liquid~\cite{Giamarchi2003}.
The ground state of the XXZ chain is the TL liquid as long as $\Delta_z$ is in the range~\eqref{Delta_z_range}, independent of the sign of $J$~\footnote{When $J>0$ and $\Delta_z=1$, the Hamiltonian \eqref{H_1d_def} is the Heisenberg antiferromagnetic chain whose ground state is the TL liquid.
However, we do not consider this case because the ESR signal of the Heisenberg chain is trivial when $\Delta_z=1$, as we discuss later.
}.
The gyromagnetic ratio $\gamma'_g$ differs from that of the ferromagnetic insulator.
$N_c$ denotes the number of spin chains that came into contact with the three-dimensional ferromagnet through the interfacial interaction $\mathcal H_{\rm int}$.
We consider that the spin chains are aligned in the same direction.
Real materials are always quasi-one-dimensional and have interchain interaction.
However, we drop them into the Hamiltonian \eqref{H_1d_def}.
This pure one-dimensional regime is realized when the temperature $T$ is much higher than the interchain exchange coupling, say $J'$: 
\begin{align}
    k_BT \gg J'.
\end{align}
In the following, we employ the unit $\hbar = k_B =1$ to lighten the notation.

Last but not least, the interfacial interaction $\mathcal H_{\rm int}$ is 
\begin{align}
    \mathcal H_{\rm int} = \sum_i \frac{J_i}2 (S_i^+s_i^-+S_i^-s_i^+),
    \label{H_int_def}
\end{align}
with $S_i^\pm=S_i^x\pm iS_i^y$ and $s_i^\pm=s_i^x\pm is_i^y$.
The coupling constant $J_i$ will depend on the site index $i$, reflecting the roughness of the interface.
To take into account the roughness, we assume that $J_i$ follows a Gaussian probability distribution.
The Gaussian distribution immediately leads to the following ensemble averages~\cite{Ominato2022a,Ominato2022b,Fukuzawa2023}:
\begin{align}
    \braket{J_i}_{\rm imp} &= \mathcal J_1, 
    \label{J_imp_av} \\
    \braket{J_i J_j}_{\rm imp} &= \mathcal J_1^2 + \mathcal J_2 \delta_{i,j},
    \label{JJ_imp_av}
\end{align}
where $\braket{\cdot}_{\rm imp}$ is the average with respect to the Gaussian probability distribution.
$\mathcal J_2$ is positive while the sign of $\mathcal J_1$ is arbitrary.

\subsection{Effective low-energy theory}

With this preparation, we proceed to deriving a low-energy effective theory by considering the interfacial interaction $\mathcal H_{\rm int}$ as a perturbation to the other two parts.
Let us adopt linear spin-wave theory to effectively describe the low-energy physics of the ferromagnetic insulator.
The linear spin-wave theory approximates the Hamiltonian as
\begin{align}
    \mathcal H_{\rm 3d} \approx \sum_{\bm k} \omega_{\bm k} b_{\bm k}^\dag b_{\bm k},
\end{align}
where $b_{\bm k}$ is the annihilation operator of the magnon with the three-dimensional wave vector $\bm k$ and 
\begin{align}
    \omega_{\bm k}=\mathcal D\bm k^2 + \gamma_g h_{\rm dc}
    \label{dispersion_NG_FI}
\end{align}
is the magnon dispersion relation.
The positive parameters $\mathcal D\propto |J_{\rm F}|S_0$ are the spin stiffness, proportional to the exchange coupling, and the magnitude $S_0$ of the local spin of the three-dimensional ferromagnet.

This paper focuses on spin pumping, triggered by the FMR of the three-dimensional ferromagnet.
Generally, FMR refers to the resonant absorption of an applied electromagnetic wave by a material in the ferromagnetic phase.
The electromagnetic wave is typically the microwave, whose long enough wavelength allows us to regard it as a plane wave.
The FMR in the three-dimensional ferromagnet is attributed to magnons with $\bm k=\bm 0$.
We thus keep the $\bm k=\bm 0$ mode and discard the others in the $b_{\bm 0}$ and $b_{\bm 0}^\dag$ are related to the raising and lowering operators $\bm S_{\bm 0}^\pm := \sum_i (S_i^x \pm i S_i^y)$ of the total spin $\bm S_{\bm 0}$.
\begin{align}
    \mathcal H_{\rm 3d} &\approx \omega_{\bm 0}b_{\bm 0}^\dag b_{\bm 0}.
\end{align}

At low temperatures, the spin-$1/2$ XXZ chain \eqref{H_1d_def} turns into the TL liquid whose Hamiltonian has the following quadratic form of a canonical pair of bosonic fields $\phi(x)$ and $\theta(x)$.
\begin{align}
    \mathcal H_{\rm 1d}/N_c \approx \frac v{2\pi} \int dx  \biggl( K(\partial_x\theta)^2 
    + \frac 1K(\partial_x \phi)^2 \biggr).
    \label{H_1d_eff}
\end{align}
Here, $K$ is the Luttinger parameter that controls the critical properties of the TL liquid, and $v$ is the spinon velocity~\cite{Giamarchi2003}.
$K$ is determined by the magnetic anisotropy and the magnetic field.
The two bosonic fields $\phi$ and $\theta$ satisfy the following commutation relation.
\begin{align}
    [\phi(x), \theta(y)] = i\pi \Theta_{\rm H}(y-x),
\end{align}
where $\Theta_{\rm H}(z)$ is the Heaviside step function,
\begin{align}
    \Theta_{\rm H}(z) &= \left\{
    \begin{array}{ccc}
        1 & & (z>0)  \\
        1/2 & & (z=0) \\
        0 & & (z<0)
    \end{array}
    \right..
    \label{Heaviside_step_def}
\end{align}

Since only the $\bm k=\bm 0$ mode is kept in the three-dimensional ferromagnet, the interfacial interaction becomes
\begin{align}
    \mathcal H_{\rm int} &\approx \sqrt{\frac{S_0}{2}} (b_{\bm 0}^\dag \tilde s^+ + b_{\bm 0} \tilde s^-), 
    \label{H_int_eff}
\end{align}
where $\tilde s^\pm$ are the following operators of the spin chain.
\begin{align}
    \tilde s^\pm := \frac{N_c}2\sum_i J_i s_i^\pm.
\end{align}
We dropped the second-order and higher-order terms about magnon operators by supposing that the magnon-magnon interaction is weak.
When approximating the interfacial interaction as in Eq.~\eqref{H_int_eff}, we take into account various $\bm k$ modes from the spin chain, while we pick up only the $\bm k=\bm 0$ mode from the three-dimensional ferromagnet.
We treated the two spin systems non-equivalently because the spin chain has low-energy excitations with wavenumbers $\bm q\approx \bm 0$ and $\bm q\approx (\pi/a_0)\bm e$, where $a_0$ is the lattice spacing of the spin chain and $\bm e$ is a unit vector parallel to the spin chains.

\section{Electron spin resonance}\label{sec:ESR}

\subsection{Introduction}

FMR is the electron spin resonance (ESR) of a material in the ferromagnetic phase.
Generally, ESR is a resonant absorption of an electromagnetic wave in the presence of a uniform static magnetic field $h_{\rm dc}$.
The applied electromagnetic wave is typically a microwave.
Let us consider a situation where we apply the monochromatic microwave with a frequency $\nu=2\pi \omega$ to our system.
Experimentally, the ESR absorption spectrum is measured by changing $h_{\rm dc}$ while fixing $\omega$.
Since changing $h_{\rm dc}$ would cause a phase transition in the target material, 
it is challenging for theoretical analyzes to deal with exactly the same situation.
Here, instead, we deal with a physically equivalent but more easily manageable problem.
We fix $h_{\rm dc}$ and change the frequency $\omega$ in this paper, as previous ESR theories do~\cite{oshikawa2002_esr,maeda2005_esrshift_jpsj,furuya2018_esr_2d}.
The linear response theory gives the following simple formula of the ESR spectrum $I(\omega)$ in the Faraday configuration~\cite{kubo-tomita1954}:
\begin{align}
    I(\omega) 
    &= -\frac{h_{\rm R}^2 \omega}{2} [\im G^R_{\mathcal S^+\mathcal S^-}(\omega) + \im G^R_{\mathcal S^-\mathcal S^+}(\omega) ],
    \label{ESR_spec_def}
\end{align}
where $h_{\rm R}$ is the magnetic-field amplitude of the applied microwave and $G^R_{\mathcal O\mathcal O^\dag}(\omega)$ is the retarded Green's function,
\begin{align}
    G^R_{\mathcal O\mathcal O^\dag}(\omega) := -i\int_0^\infty dt \, e^{i\omega t} \braket{[\mathcal O(t), \mathcal O^\dag(0)]}.
\end{align}
The operators $\mathcal S^+$ and $\mathcal S^-=(\mathcal S^+)^\dag$ are the raising and lowering operators of the total spin of the system,
\begin{align}
    \mathcal S^\pm = \sum_i S_i^+ + \sum_i s_i^+.
    \label{total_ladder_op_def}
\end{align}
The expression \eqref{ESR_spec_def} implicitly assumes that the applied microwave is unpolarized~\cite{furuya2018_esr_2d}.
Since the interface coupling $J_i$ is much weaker than $J_{\rm F}$ and $J$, the ESR spectrum \eqref{ESR_spec_def} can be approximated as a simple superposition of those isolated from the bulk ferromagnetic insulator and the spin chain compound, that is,
\begin{align}
    I(\omega)
    &\approx I_{\rm FI}(\omega) + I_{\rm 1d}(\omega),  
    \label{ESR_spec_ours_def} \\
    I_{\rm FI}(\omega)
    &\approx -\frac{h_{\rm R}^2\omega}{2} [\im G^R_{S^+S^-}(\omega) +\im G^R_{S^-S^+}(\omega) ],
    \label{ESR_spec_FI_def} \\
    I_{\rm 1d}(\omega)
    &= - \frac{h_{\rm R}^2\omega}{2} [ \im G^R_{s^+s^-}(\omega) + \im G^R_{s^-s^+}(\omega)
    ].
    \label{ESR_spec_1d_def}
\end{align}
$I_{\rm FI}(\omega)$ represents the FMR in the three-dimensional ferromagnet and $I_{\rm 1d}(\omega)$ represents the ESR in the spin-chain material.

\subsection{ESR Linewidth of quantum antiferromagnetic spin chain}\label{sec:review_esr}

Let us first review an ESR theory of the quantum antiferromagnetic spin chain~\cite{oshikawa2002_esr}.
If $\Delta_z$ exactly equals to 1, the ESR spectrum \eqref{ESR_spec_1d_def} due to the spin chain would be trivially given by the delta function~\cite{oshikawa2002_esr}:
\begin{align}
    I_{\rm 1d}(\omega)
    &= \pi h_{\rm R}^2\omega\braket{s_{\bm 0}^z} \delta(\omega-\gamma'_g h_{\rm dc}),
    \label{I_1d_isotropic}
\end{align}
where $\braket{s_{\bm 0}^z}= \sum_i\braket{s_i^z}$ is the total magnetization of the spin-chain material.
In the classical-spin picture, the delta function of the ESR spectrum means that the total spin $\bm s_{\bm 0}=\sum_i \bm s_i$ around the static magnetic field precesses forever without any disturbance.

Magnetic anisotropy with $\Delta_z\not=1$ disturbs the precession of the total spin and gives the finite linewidth to the ESR absorption peak of $I_{\rm 1d}(\omega)$.
An explicit analytic representation of the retarded Green's function $G^R_{s^\pm s^\mp}(\omega)$ is available in the TL-liquid phase.
For the nearly isotropic \emph{antiferromagnetic} XXZ chain ($J>0$ and $0< 1- \Delta_z \ll 1$),
we obtain~\cite{oshikawa2002_esr}
\begin{align}
    \im G^R_{s^+s^-}(\omega)
    &\approx -\frac{8\pi^2 \gamma'_g h_{\rm dc} \Delta'T/v}{(\omega-\gamma'_g h_{\rm dc})^2+(4\pi \Delta'T)^2}.
\end{align}
This imaginary part approaches the delta function in the isotropic limit, $\Delta_z \to 1-0$.
\begin{align}
    \lim_{\Delta_z \to 1-0} [-\im G^R_{s^-s^+}(\omega)] \propto  \frac{2\pi \gamma'_g h_{\rm dc}}v \delta(\omega - \gamma'_g h_{\rm dc}).
\end{align}
The ESR absorption peak of the spin chain thus has a Lorentzian lineshape with the linewidth proportional to $(1-\Delta_z)^2T$.

\subsection{Gilbert damping}

Let us move on to the FMR spectrum \eqref{ESR_spec_FI_def}.
If the interfacial interaction is absent, the FMR spectrum is easily obtained since the Hamiltonian \eqref{H_FI_def} of the ferromagnetic insulator exactly gives
\begin{align}
   \im G_{S^+S^-}^R(\omega) &= - 2\pi \braket{S_{\bm 0}^z} \delta(\omega-\gamma_g h_{\rm dc}),
   \label{GS+S-_FI_isotropic} \\
   \im G_{S^-S^+}^R(\omega) &= 2\pi \braket{S_{\bm 0}^z} \delta(\omega+\gamma_g h_{\rm dc})
   \label{GS-S+_FI_isotropic} \\
   &= 0,
\end{align}
in exactly the same fashion with the ESR spectrum \eqref{I_1d_isotropic} for the Heisenberg chain under the magnetic field.
Note that $\omega +\gamma h_{\rm dc}>0$ for $h_{\rm dc}>0$.
Since the isotropic Heisenberg interaction does not affect the dynamics of the total spin $\bm S_{\bm 0}$, the total spin precesses around the direction of the magnetic field forever without any damping.
This damping-free precession gives the sharp delta-function peak with zero linewidth.

The interfacial interaction \eqref{H_int_eff} introduces damping effects on the FMR spectrum \eqref{ESR_spec_FI_def}.
We adopt a phenomenological approach to dealing with them.
The total spin $\bm S_{\bm 0}$ in the ferromagnetic phase is often well described by the following phenomenological equation of motion.
\begin{align}
    \frac{d\bm S_{\bm 0}}{dt}
    &= -\gamma_g \bm h_{\rm dc} \times \bm S_{\bm 0} - \frac{\alpha_{\rm G}}{S_0} \bm S_{\bm 0} \times \frac{d\bm S_{\bm 0}}{dt},
    \label{LLG_def}
\end{align}
where $\alpha_{\rm G}>0$ is the Gilbert damping coefficient and $\bm h_{\rm dc} = h_{\rm dc} \bm e_z$.
Equation \eqref{LLG_def} is called the Landau-Lifshitz-Gilbert (LLG) equation.
The second term of the LLG equation \eqref{LLG_def} represents the damping effect of the precession of $\bm S_{\bm 0}$.
Should $\alpha_{\rm G}$ be zero, Eq.~\eqref{LLG_def} claims that the precession of $\bm S_{\bm 0}$ around $\bm h_{\rm dc}$ lasts forever, as represented by Eqs.~\eqref{GS+S-_FI_isotropic} and \eqref{GS-S+_FI_isotropic}.

When $\bm S_{\bm 0}$ follows the LLG equation \eqref{LLG_def}, their Green's functions $G^R_{S^+S^-}(\omega)$ and $G^R_{S^-S^+}(\omega)$ acquire the finite imaginary part.
\begin{align}
    G^R_{S^+S^-}(\omega) &\approx \frac{2S_0 \braket{b_{\bm 0}b_{\bm 0}^\dag}}{\omega-\gamma_g h_{\rm dc} + i\alpha_{\rm G}\omega},
    \label{GR_+_LLG} \\
    G^R_{S^-S^+}(\omega) &\approx \frac{2S_0\braket{b_{\bm 0}^\dag b_{\bm 0}}}{\omega +\gamma_g h_{\rm dc}-i\alpha_{\rm G}\omega}.
    \label{GR_-_LLG}
\end{align}
Equations~\eqref{GR_+_LLG} and \eqref{GR_-_LLG} hold in the ferromagnetically ordered phase at low temperatures so that the magnon density satisfies $\braket{b_{\bm 0}^\dag b_{\bm 0}} \ll 2S_0$.
These Green's functions are derived from the LLG equation in Appendix~\ref{app:LLG_to_GR}.
The FMR occurs around $\omega= \gamma_g h_{\rm dc}$.
Note that the FMR around $\omega = \gamma_g h_{\rm dc}$ is attributed to $G^R_{S^+S^-}(\omega)$ while the FMR around $\omega = -\gamma_g h_{\rm dc}$ is to $G^R_{S^-S^+}(\omega)$.
In the following, we focus on $G^R_{S^+S^-}(\omega)$ and ignore $G^R_{S^-S^+}(\omega)$ since $h_{\rm dc}>0$.

The interfacial interaction is not the only source of Gilbert damping.
Many interactions (not included in our model) can potentially induce the damping effect in the LLG equation~\eqref{LLG_def}.
A representative is the magnon-phonon interaction.
Suppose that the three-dimensional ferromagnetic insulator has the Gilbert damping $\alpha_{\rm G}>0$ in the absence of interfacial interaction.
The activation of the interfacial interaction changes the Gilbert damping coefficient from $\alpha_{\rm G}$ to $\alpha_{\rm G}+\delta \alpha_{\rm G}$.
The increase in Gilbert damping due to the interfacial interaction $\delta \alpha_{\rm G}$ is perturbatively given by the following self-energy~\cite{Fukuzawa2023}
\begin{align}
    \delta \alpha_{\rm G}
    &= - \frac{2S_0}{\gamma_g h_{\rm dc}} \im \Sigma^R(\gamma_g h_{\rm dc}).
    \label{alpha_self-energy}
\end{align}
$\Sigma^R(\omega)$ is the Fourier transform of the retarded Green's function,
\begin{align}
    \Sigma^R(\omega)
    &= \int_{-\infty}^\infty dt \, \Sigma^R(t),
    \label{Self-energy_omega_def} \\
    \Sigma^R(t) &= -i\theta_{\rm H}(t) \braket{\braket{[\tilde s^+(t), \tilde s^-(0)]}}_{\rm imp}.
    \label{Self-energy_t_def}
\end{align}
Here, $\Theta_{\rm H}(t)$ is the Heaviside's step function.
Taking the random average [see Eqs.~\eqref{J_imp_av} and \eqref{JJ_imp_av}] about $J_i$, we obtain
\begin{align}
    \Sigma^R(t)
    &= \Sigma_1^R(t) + \Sigma_2^R(t),
    \label{Self-energy_12_def} \\
    \Sigma_1^R(t)
    &= -\frac i4 N_c \mathcal J_1^2  \Theta_{\rm H}(t) \sum_{j, k} \braket{[s_{j}^+(t), s_{k}^-(0)]},
    \label{Self-energy_1_def} \\
    \Sigma_2^R(t)
    &= - \frac i4 N_c\mathcal J_2 \Theta_{\rm H}(t) \sum_{j} \braket{[s_{j}^+(t), s_{j}^-(0)]}.
    \label{Self-energy_2_def}
\end{align}
$\Sigma_1^R(t)$ reads as the uniform correlation over the spin chain while $\Sigma_2^R(t)$ as the local one.
The parameter $\mathcal J_2$ measures the strength of randomness.
The interface is clean for $\mathcal J_2=0$ while dirty for $\mathcal J_2 \gtrsim |\mathcal J_1|$.
Let us call $\Sigma_1^R(t)$ the clean part and $\Sigma_2^R(t)$ the dirty part of the self-energy.
Likewise, we split the Fourier transform $\Sigma^R(\omega)$ in the clean ($n=1$) and dirty ($n=2$) parts:
\begin{align}
    \Sigma_n^R(\omega) = \int_{-\infty}^\infty dt \, e^{i\omega t} \Sigma_n^R(t).
\end{align}

Note that the relation \eqref{alpha_self-energy} holds when the interfacial interaction $\mathcal H_{\rm int}$ is a perturbation to the other parts of the total Hamiltonian \eqref{H_tot_def}~\cite{Fukuzawa2023}.
Hence, the correlations in the self energies \eqref{Self-energy_1_def} and \eqref{Self-energy_2_def} are evaluated as those of the single isolated spin chain.
Such calculations are straightforwardly performed with the aid of two-dimensional conformal field theory~\cite{Voit1995,vonDelft1998,Giamarchi2003}.
In the subsequent sections, we discuss the clean and dirty parts of the self-energy based on the conformal field theory.

\section{Ferromagnetic chains}\label{sec:FM}

\subsection{TL liquid with large Luttinger parameter}\label{sec:TL_large-K}

Let us deal with \emph{ferromagnetic} XXZ chain with $J<0$ and small magnetic anisotropy $0< 1-\Delta_z\ll 1$.
As long as the anisotropy $\Delta_z$ is in the range \eqref{Delta_z_range}, the ground state is the TL liquid with no long-range orders.
We focus on the ferromagnetic XXZ chain for the following reason.
The previous study~\cite{Fukuzawa2023} in a similar setup shows that the Gilbert damping coefficient $\alpha_{\rm G}$ exhibits a power-law dependence on temperature.
The power is governed by the Luttinger parameter of the TL liquid.
In our situation with $J<0$, the Luttinger parameter $K$ diverges as $\Delta_z \to 1-0$ due to the quantum phase transition from the TL liquid to the ferromagnetic phase at $\Delta_z=1$.
When $J<0$ and $h_{\rm dc}=0$, the Luttinger parameter $K$ shows~\cite{cabra_ladder,Giamarchi2003}
\begin{align}
    K = \frac{\pi}{2\cos^{-1}\Delta_z}.
    \label{K_large}
\end{align}
The TL liquid with such a large Luttinger parameter was hardly discussed thus far.
We show characteristic behavior of the TL liquid with the large Luttinger parameter in spin pumping in the rest of this paper.

\subsection{Bosonization}\label{sec:bosonization}

The ferromagnetic XXZ chain \eqref{H_1d_def} with a small magnetic anisotropy is equivalent to an antiferromagnetic XXZ chain with a large magnetic anisotropy.
We can see this equivalence by performing the following staggered rotation,
\begin{align}
    s'^z_j = s_j^z, \quad s'^\pm_\pm =  (-1)^j s_j^\pm.
    \label{pi_rot_sz}
\end{align}
Then, the Hamiltonian \eqref{H_1d_def} becomes
\begin{align}
    \mathcal H_{\rm 1d}/N_c\
    &=|J| \sum_j (s'^x_js'^x_{j+1} + s'^y_j s'^y_{j+1} - \Delta_z s'^z_j s'^z_{j+1}) 
    \notag \\
    &\qquad - \gamma'_g h_{\rm dc} \sum_j s'^z_j.
    \label{H_1d_pi_rotated}
\end{align}

At low temperatures $T\ll J$, the antiferromagnetic XXZ chain \eqref{H_1d_pi_rotated} turns into the TL-liquid Hamiltonian \eqref{H_1d_eff}.
This transformation relates the spin operator $\bm s'_j$ and the boson fields $\phi$ and $\theta$ as~\cite{Giamarchi2003}
\begin{align}
    s'^z_j &= m + \frac{a_0}{\pi}\partial_x \phi + (-1)^j a_1 \sin(2\phi + 2\pi mx/a_0), 
    \label{sz_bosonization_afm} \\
    s'^+_j &= e^{-i\theta}[ b_0 (-1)^j + b_1 \sin (2\phi + 2\pi mx/a_0)], 
    \label{s+_bosonization_afm} \\
    s'^-_j &= (s'^+_j)^\dag.
\end{align}
Here, $m$ is the magnetization per site along the field and $a_1$, $b_0$, and $b_1$ are non-universal parameters~\cite{hikihara2004_amplitude_bosonization}.
Back to the original ferromagnetic XXZ chain, we obtain
\begin{align}
    s^z_j &= m + \frac{a_0}{\pi}\partial_x \phi + (-1)^j a_1 \sin(2\phi + 2\pi mx/a_0), 
    \label{sz_bosonization_fm} \\
    s^+_j &= e^{-i\theta}[ b_0 + b_1 (-1)^j \sin (2\phi + 2\pi mx/a_0)], 
    \label{s+_bosonization_fm} \\
    s^-_j &= (s'^+_j)^\dag.
\end{align}
When $\Delta_z \approx 1$, the operator $e^{-i\theta}$ with the scaling dimension $1/4K \ll 1$ is much more relevant than $\sin(2\phi + 2\pi mx/a_0)$ with the scaling dimension $K\gg 1$.
Hence, we can approximate $s_j^\pm \approx b_0 e^{\mp i\theta}$ by calculating their retarded Green's functions.

\subsection{Clean part}\label{sec:clean}

The clean part \eqref{Self-energy_1_def} of the self energy
\begin{align}
    \Sigma_1^R(t)
    &\approx -\frac{i}{4}N_c \mathcal J_1^2 \Theta_{\rm H}(t) b_0^2 \int_0^W dx  \int_0^W dy \braket{[e^{-i\theta(t,x)}, e^{i\theta(0,y)}]}
\end{align}
can be calculated with the aid of the conformal symmetry~\cite{Fukuzawa2023}.
Its Fourier transform is given by 
\begin{align}
    \Sigma_1^R(\omega)
    &=\frac{N_c\mathcal J_1^2 b_0^2}{2} \int_0^\infty dt \, e^{i\omega t} \int_0^W dx \int_0^W dy \, \im g(t,x-y),
    \label{Sigma_1_clean_integrals}
\end{align}
where $g(t,z)$ is the following function.
\begin{align}
    g(t,z)
    &= \biggl(\frac{\sinh(i\pi a_0/\beta v)}{\sinh[\pi(ia_0-z-vt)/\beta v]}\biggr)^{\frac{1}{4K}}
    \notag \\
    &\qquad \times \biggl(\frac{\sinh(i\pi a_0/\beta v)}{\sinh[\pi (ia_0+z-vt)/\beta v]}
    \biggr)^{\frac{1}{4K}}
    \label{g_def}
\end{align}
In what follows, we evaluate the self energy \eqref{Sigma_1_clean_integrals} in two opposite limits, a short-junction limit ($W\ll \ell_{\rm th}$) and a long-junction limit ($W \gg \ell_{\rm th}$), where
\begin{align}
    \ell_{\rm th} = \frac{v}{\pi T}.
    \label{thermal_length}
\end{align}
The length scale $\ell_{\rm th}$ gives the characteristic decay length of the exponentially decaying $g(t,z)$ as $|z|$ increases.
We call $\ell_{\rm th}$ thermal length.

\subsubsection{Short-junction limit}

Since $g(t,z)$ decays exponentially and vanishes for $|z|\gtrsim \ell_{\rm th}$ and $\ell_{\rm th}\gg W$ in the short-junction limit, we may approximate $g(t,x-y)\approx g(t,0)$ in the integrand, leading to a significant simplification of calculations.
The details of the calculations are shown in Appendix~\ref{app:clean_short}.
The final result for $\im \Sigma_1^R(\omega)$ is
\begin{align}
    \im \Sigma_1^R(\omega) &\approx -\frac{\pi N_c\mathcal J_1^2b_0^2W^2}{4v^2} \omega \biggl(\frac{2\pi a_0}{\beta v}\biggr)^{\frac 1{2K}-2}\frac{\Gamma(\frac{1}{4K})^2}{\Gamma(\frac{1}{2K})}.
    \label{self-energy_clean_short}
\end{align}
When $K\gg 1$, this expression is further simplified as
\begin{align}
    \im \Sigma_1^R(\omega)
    &\approx -\frac{2\pi N_c\mathcal J_1^2b_0^2W^2K}{v^2} \omega \biggl(\frac{2\pi a_0}{\beta v}\biggr)^{\frac 1{2K}-2},
    \label{self-energy_clean_short_large_K}
\end{align}
leading to
\begin{align}
    \delta \alpha_{\rm G,1}
    &\approx \frac{4\pi S_{\bm 0} N_c\mathcal J_1^2b_0^2W^2K}{v^2} \biggl(\frac{2\pi a_0}{\beta v}\biggr)^{\frac 1{2K}-2}.
    \label{alpha_clean_short_large_K}
\end{align}
Note that we made an assumption of $\omega/T \ll 1$ to derive Eq.~\eqref{self-energy_clean_short}.
Since $\omega=\gamma_g h_{\rm dc}$ in our spin pumping, we can rephrase the assumption as $\gamma_g h_{\rm dc}/T \ll 1$, which holds at low magnetic fields.
The power $\delta \alpha_{\rm G,1} \propto (T/v)^{\frac 1{2K}-2}$ depends on the Luttinger parameter $K$, which is the main feature of the TL liquid.
Despite the criticality, the power is not universal, but depends on the non-universal parameter $K$.
As given in \eqref{K_large}, the Luttinger parameter $K$ is extremely large, as we have already mentioned.
The large negative power $1/2K- 2 \approx - 2$ leads to strong damping in the spin pumping experiment at low temperatures.

\subsubsection{Long-junction limit}

The exponentially decaying function $g(t,z)$ is almost zero for $|z|\gg \ell_{\rm th}$.
Hence, we may replace $W\to +\infty$ in the spatial integration range.
Then the problem is reduced to the textbook calculation of the retarded Green's function (see Appendix C of Ref.~\cite{Giamarchi2003}).
The clean part of the self energy then becomes
\begin{align}
    \Sigma_1^R(\omega)
    &= - \frac{N_c\mathcal J_1^2b_0^2W}{4a_0^2v} \biggl(\frac{2\pi a_0}{\beta v}\biggr)^{\frac{1}{2K}} \sin\biggl(\frac{\pi}{4K}\biggr) \biggl(\frac{\beta v}{2\pi}\biggr)^2 
    \notag \\
    &\qquad \times \biggl[
    B\biggl(\frac{1}{8K}-i\frac{\beta\omega}{4\pi}, 1-\frac{1}{4K}\biggr)
    \biggr]^2.
    \label{self-energy_clean_long_full}
\end{align}
The self energy at $\omega=\gamma_g h_{\rm dc}$ is relevant to the Gilbert damping \eqref{alpha_self-energy}.
When $\omega \ll T$, the clean part \eqref{self-energy_clean_long_full} is simplified as
\begin{align}
    \im \Sigma_1^R(\omega) &\approx  -\frac{N_c\mathcal J_1^2b_0^2 W}{16v} \beta\omega \biggl(\frac{2\pi a_0}{\beta v}\biggr)^{\frac{1}{2K}-2}  \biggl(\frac{\Gamma(\frac{1}{8K})}{\Gamma(\frac{1}{4K})}\biggr)^2 .
    \label{self-energy_clean_long}
\end{align}
For $K\gg 1$, this expression is furthermore simplified.
\begin{align}
    \im \Sigma_1^R(\omega) &\approx - \frac{N_c\mathcal J_1^2b_0^2WK}{\pi v} \beta \omega \biggl(\frac{2\pi a_0}{\beta v}\biggr)^{\frac{1}{2K}-2}.
    \label{self-energy_clean_long_large_K}
\end{align}
The Gilbert damping coefficient then becomes
\begin{align}
    \delta\alpha_{\rm G,1}
    &\approx \frac{2S_{\bm 0}N_c\mathcal J_1^2b_0^2WK}{\pi v} \beta \biggl(\frac{2\pi a_0}{\beta v}\biggr)^{\frac{1}{2K}-2}.
    \label{alpha_clean_long_large_K}
\end{align}
The power law $\delta \alpha_{\rm G,1} \propto (T/v)^{\frac 1{2K}-3}$ of Eq.~\eqref{alpha_clean_long_large_K} on the temperature differs from  Eq.~\eqref{alpha_clean_short_large_K} in the short-junction limit.
However, we found that $\delta\alpha_{\rm G,1}$
is strongly enhanced at low temperatures $T/v\ll 1$ regardless of the length $W$ of the junction.
This enhancement originates from the $\omega\to +0$ limit of the $q=0$ component of the transverse correlation $\braket{s^+s^-}(\omega,q)$.
Taking the $\omega \to +0$ limit, the transverse correlation function $\braket{s^+s^-}(\omega,0)$ diverges since the ferromagnetic correlation at $q\approx 0$ is well developed in our ferromagnetic chains even though the easy-plane magnetic anisotropy is present.

\subsection{Dirty part}\label{sec:dirty}

The dirty part \eqref{Self-energy_2_def} of the self-energy
\begin{align}
    \Sigma_2^R(t)
    &= -\frac i4 N_c \mathcal J_2\Theta_{\rm H}(t) b_0^2 \int_0^W dx \, \braket{[e^{-i\theta(t,x)}, e^{i\theta(0,x)}]}
\end{align}
is also related to $g(t,z)$ of Eq.~\eqref{g_def}.
The Fourier transform is given by
\begin{align}
    \Sigma_2^R(\omega) 
    &\approx - i\frac{\pi N_c \mathcal J_2b_0^2a_0 W}{2v^2} \omega \biggl(\frac{2\pi a_0}{\beta v}\biggr)^{\frac 1{2K}-2} \frac{\Gamma(\frac{1}{4K})^2}{\Gamma(\frac{1}{2K})}.
        \label{self-energy_dirty}
\end{align}
for any $W$.
In the large-$K$ limit, we can approximate it as
\begin{align}
    \Sigma_2^R(\omega) &\approx -i\frac{4\pi N_c\mathcal J_2b_0^2 a_0WK}{v^2} \omega \biggl(\frac{2\pi a_0}{\beta v}\biggr)^{\frac{1}{2K}-2}.
    \label{self-energy_dirty_large_K}
\end{align}
We thus reach the following representation.
\begin{align}
    \delta\alpha_{\rm G,2}
    &\approx \frac{8\pi S_{\bm 0}N_c\mathcal J_2b_0^2 a_0WK}{v^2}  \biggl(\frac{2\pi a_0}{\beta v}\biggr)^{\frac{1}{2K}-2}.
    \label{alpha_dirty_large_K}
\end{align}
We obtain the power law $\delta\alpha_{\rm G,2} \propto (T/v)^{\frac 1{2K}-2}$ same as that of Eq.~\eqref{alpha_clean_short_large_K} in the short-junction limit.
Whereas only the $q\approx 0$ mode can contribute to the clean part, any $q$ modes can potentially contribute equally to the dirty part.
The same power law means that the $q\approx 0$ mode still stands out even if any other modes are taken into account.
We find this result quite natural, considering the fact that our system is located near the quantum critical point to the ferromagnetically ordered phase.

\subsection{Gilbert damping}

We plug in the self energies into the formula \eqref{alpha_self-energy} and obtain
\begin{align}
    \delta \alpha_{\rm G} &= \delta \alpha_{\rm G,1} + \delta \alpha_{\rm G,2}.
\end{align}
As we already saw, each term shows the following temperature dependence.
\begin{align}
    \delta \alpha_{\rm G,1}
    &\propto \left\{
    \begin{array}{ccc}
      \mathcal J_1^2W^2K T^{\frac{1}{2K}-2}  & & (W \ll \ell_{\rm th}), \\[+10pt]
      \mathcal J_1^2WK T^{\frac{1}{2K}-3} & & (W \gg \ell_{\rm th}),
    \end{array}
    \right.
    \label{alpha_G1_power}
\end{align}
and
\begin{align}
    \delta \alpha_{\rm G,2}
    &\propto \mathcal J_2 WK T^{\frac{1}{2K}-2}.
    \label{alpha_G2_power}
\end{align}

\begin{figure}[t!]
    \centering
    \includegraphics[bb = 0 0 1600 600, width=\linewidth]{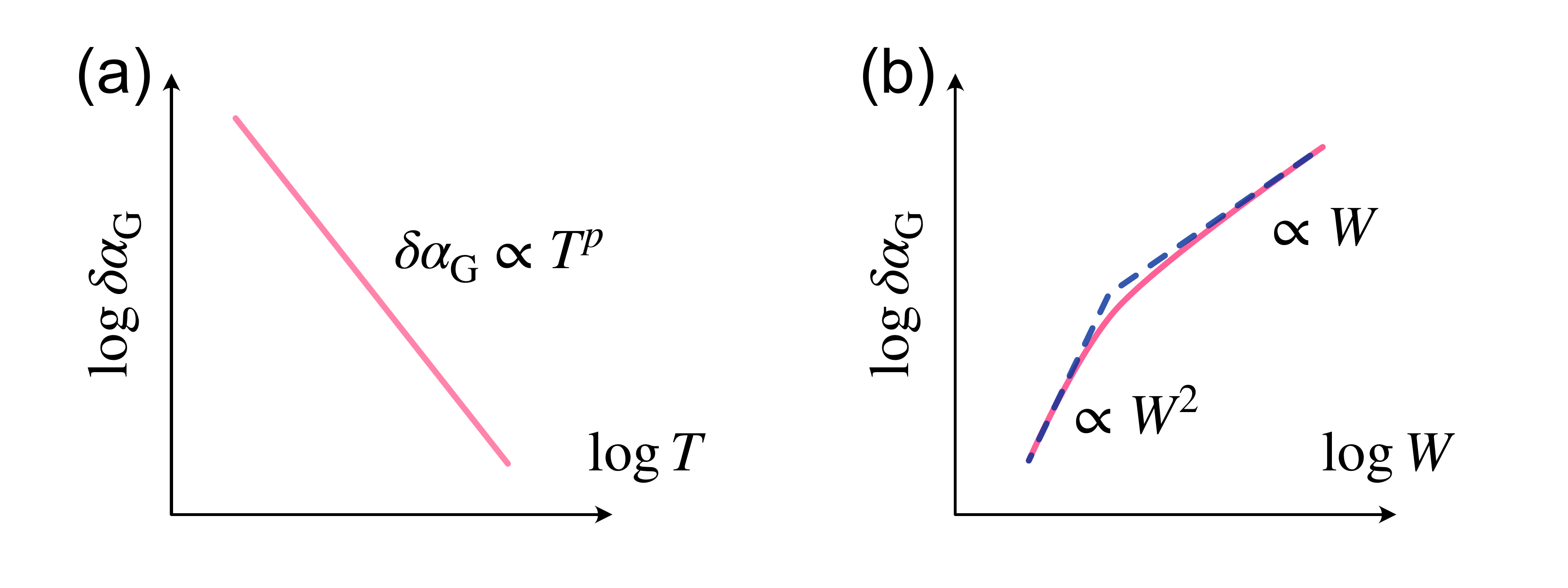}
    \caption{Schematic figures of parameter dependence of $\delta \alpha_{\rm G}$ in the clean limit. (a) Temperature dependence. The Gilbert damping $\delta \alpha_{\rm G}$ shows a power-law dependence on the temperature $T$. The power $p$ is either $\frac{1}{2K}-2$ or $\frac{1}{2K}-3$, depending on the length $W$ of the junction [Eq.~\eqref{alpha_G1_power}]. (b) Length dependence of $\delta \alpha_{\rm G}$. 
    The Gilbert damping also shows a power-law dependence on the junction length $W$. 
    The power changes roughly at the thermal length $\ell_{\rm th}$ the  when the junction is clean enough:
    $\delta \alpha_{\rm G}$ for $W \ll \ell_{\rm th}$ and $\delta \alpha_{\rm G} \propto W$ for $W\gg \ell_{\rm th}$ [Eq.~\eqref{alpha_G1_power}].
    }
    \label{fig:alpha}
\end{figure}

When the interface is rough such that $\mathcal J_2a_0 \gg \mathcal J_1^2 W$,
the dirty part dominates the Gilbert damping (\textit{i.e.}, $\delta\alpha_{\rm G,2} \gg \delta \alpha_{\rm G,1})$ only in the short-junction limit.
Despite the roughness of the interface, the clean part is the main source of damping in the long-junction limit due to the large negative power dependence of $\delta \alpha_{\rm G,1}$ on the temperature $T$.
Thus, we end up with
\begin{align}
    \delta \alpha_{\rm G}
    &\approx \left\{
    \begin{array}{ccc}
        \delta \alpha_{\rm G,2} & & (W \ll \ell_{\rm th},\quad \mathcal J_2a_0 \gg \mathcal J_1^2 W), \\[+10pt]
        \delta \alpha_{\rm G,1} & & (\ell_{\rm th}\ll W, \quad \mathcal J_2a_0 \gg \mathcal J_1^2 W).
    \end{array}
    \right.
    \label{total_alpha_rough_final}
\end{align}
By contrast, when the interface is clean so that $\mathcal J_2a_0\ll \mathcal J_1^2 W$, the clean part always wins.
\begin{align}
    \delta \alpha_{\rm G } &\approx \delta\alpha_{\rm G,1}, \quad (\mathcal J_2a_0\ll \mathcal J_1^2 W).
    \label{total_alpha_clean_final}
\end{align}
Figure~\ref{fig:alpha} shows the temperature and junction-length dependence of the Gilbert damping in the clean case \eqref{total_alpha_clean_final}.

The Gilbert damping in our system qualitatively and quantitatively differs from that investigated in the previous study~\cite{Fukuzawa2023} for the carbon nanotube.
The power law of the temperature dependence differs quantitatively.
The powers $\frac 1{2K}-2$ or $\frac{1}{2K}-3$ of Eqs.~\eqref{alpha_G1_power} and \eqref{alpha_G2_power} are roughly $-2$ or $-3$ since $K\gg 1$.
On the other hand, the corresponding powers $2\gamma-2$ and $2\gamma-3$, depending on the roughness and length $W$ of the interface, of Ref.~\cite{Fukuzawa2023} is roughly $0$ or $-1$ since the parameter $\gamma \approx 1$.
This difference in power law originates from the large Luttinger parameter $K$ in our spin chains.
The weakly anisotropic ferromagnetic chain has the large $K\gg 1$, while the carbon nanotube corresponds to $K\approx 1$.
The TL liquid with $K\approx 1$ is equivalent to a weakly interacting fermion chain, but the one with $K\gg 1$ is equivalent to a strongly interacting fermion chain~\cite{Giamarchi2003}.

\section{Toward experimental realization}\label{sec:experiment}

\subsection{Candidate material}\label{sec:material}

Let us discuss the experimental feasibility of our model.
The choice of spin-chain material is the most non-trivial part of our system toward experimental realization.
Here we propose that CsCuCl$_3$ will be suitable for this purpose for the following reasons.
The magnetic properties of CsCuCl$_3$ are described at low temperatures by the following quasi-one-dimensional model~\cite{Tanaka1992_CsCuCl3_ESR,Nihongi2022_CsCuCl3}.
\begin{align}
    &\mathcal H_{\rm CsCuCl_3}
    \notag \\
    &= -2J_0\sum_{j,\mu} (s_{j,\mu}^x s_{j+1,\mu}^x + s_{j,\mu}^y s_{j+1,\mu}^y + \Delta_z s_{j,\mu}^z s_{j+1,\mu}^z)
    \notag \\
    &\qquad 
    + 2J_1 \sum_{j,\mu} \bm s_{j,\mu} \cdot (\bm s_{j,\mu+1}+\bm s_{j+1,\mu+1})
    \notag \\
    &\qquad - \sum_{j,\mu} \bm D \cdot \bm s_{j,\mu} \times \bm s_{j+1,\mu}
    \notag \\
    &\qquad - \gamma_g h_{\rm dc} \sum_{j,\mu} s_{j,\mu}^z,
    \label{H_CsCuCl3}
\end{align}
where $2J_0\approx 56$~K and $2J_1 \approx 9.8$~K~\cite{Tanaka1992_CsCuCl3_ESR} are the ferromagnetic intrachain and antiferromagnetic interchain exchange couplings, respectively.
$0<\Delta_z<1$ is the easy-plane anisotropy.
The uniform DM interaction has the DM vector $\bm D=D\bm e_z$ parallel to the chain direction, which we define as the $z$ axis.
$D$ is much smaller than $2J_0$ ($D\approx 5.1$~K~\cite{Tanaka1992_CsCuCl3_ESR}).
The precise value of $\Delta_z$ is not known, but is believed to be close to 1 (i.e., $0<1-\Delta_z \ll 1$).
Thus, CsCuCl$_3$ has weak magnetic anisotropies.
We apply a uniform static magnetic field along the chain direction.

The Hamiltonian \eqref{H_CsCuCl3} contains two ingredients that have not been considered so far: the interchain interaction and the uniform DM interaction.
In what follows, we show that these two interactions have little effect on the spin pumping in a certain temperature range.
The interchain coupling is much smaller than the intrachain one, $J_1/J_0\approx 0.18$.
The interchain interaction leads to the spontaneous antiferromagnetic order.
The N\'eel temperature is experimentally determined as $T_N \approx 10.7$~K~\cite{Rioux1969_CsCuCl3,Adachi1980_CuCsCl3,Hyodo1981_CsCuCl3}.
The interchain interaction $J_1$ is negligible when the temperature $T$ is well above $T_N$.

We can eliminate the DM interaction by performing a spin rotation.
We ignore the interchain interaction and consider the following simplified Hamiltonian of CsCuCl$_3$.
\begin{align}
    &\mathcal H_{\rm CsCuCl3}
    \notag \\
    &\approx -2J_0\sum_{j,\mu} (s_{j,\mu}^x s_{j+1,\mu}^x + s_{j,\mu}^y s_{j+1,\mu}^y + \Delta_z s_{j,\mu}^z s_{j+1,\mu}^z)
    \notag \\
    &\qquad - \sum_{j,\mu} D (s_{j,\mu}^x s_{j+1,\mu}^y - s_{j,\mu}^y s_{j+1,\mu}^x)
    \notag \\
    &\qquad - \gamma_g h_{\rm dc} \sum_{j} s_{j,\mu}^z.
    \label{H_CsCuCl3_1d}
\end{align}
The gauge transformation is given by the following rotation around the $s^z$ axis:
\begin{align}
    \left\{
    \begin{array}{l}
        s_{j,\mu}^\pm  = e^{\pm i\theta_D j}\tilde s_{j,\mu}^\pm, \\ 
        \\
        s_{j,\mu}^z = \tilde s_{j,\mu}^z.        
    \end{array}
    \right.
    \label{rot_DM}
\end{align}
Here, the real parameter $\theta_D$ is given by
\begin{align}
    \theta_D  &= \tan^{-1} \biggl(\frac{D}{2J_0}\biggr).
    \label{theta_D}
\end{align}
The Hamiltonian in the rotated framework reads as
\begin{align}
    &\mathcal H_{\rm CsCuCl3}
    \notag \\
    &\approx -2\tilde J_0 \sum_{j,\mu} (\tilde s_{j,\mu}^x \tilde s_{j+1,\mu}^x + \tilde s_{j,\mu}^y \tilde s_{j+1,\mu}^y + \tilde \Delta_z \tilde s_{j,\mu}^z \tilde s_{j+1,\mu}^z)
    \notag \\
    &\qquad - \gamma_g h_{\rm dc} \sum_{j} \tilde s_{j,\mu}^z,
    \label{H_CsCuCl3_1d_rotated}
\end{align}
where
\begin{align}
    2\tilde J_0 &= \sqrt{(2J_0)^2 + D^2},
    \label{tilde_J0} \\
    \tilde \Delta_z &= \frac{J_0}{\tilde J_0} \Delta_z.
    \label{tilde_Delta_z}
\end{align}
The Hamiltonian \eqref{H_CsCuCl3_1d_rotated} in the rotated coordinate $(\tilde s_{j,\mu}^x,\tilde s_{j,\mu}^y,\tilde s_{j,\mu}^z)$ equals to the one we have investigated in the previous sections.

The rotation \eqref{rot_DM} modifies the definitions of self-energies \eqref{Self-energy_1_def} and \eqref{Self-energy_2_def}.
\begin{align}
    \Sigma_1^R(t) 
    &= -\frac i4 N_c \mathcal J_1^2 \Theta_{\rm H}(t) \sum_{j,k} e^{i\theta_D (j-k)} \braket{[\tilde s_j^+(t), \tilde s_k^-(0)]}, \\
    \Sigma_2^R(t)
    &= -\frac i4 N_c \mathcal J_2 \Theta_{\rm H}(t) \sum_{j} \braket{[\tilde s_j^+(t), \tilde s_j^-(0)]}.
\end{align}
The dirty part $\Sigma_2^R(t)$ is kept intact, as the right-hand side is the on-site correlation, independent of the position-dependent spin rotation \eqref{rot_DM}.
By contrast, the clean part is modified so that the Green's function on the right-hand side is to be evaluated at the wavenumber $k=\theta_D$ instead of $k=0$ as we did in Sec.~\ref{sec:clean}.
This modulation of the wavenumber affects the clean part of the Gilbert damping in the long-junction limit, changing Eq.~\eqref{self-energy_clean_long_full} to
\begin{align}
    \Sigma_1^R(\omega)
    &= - \frac{N_c \mathcal J_1^2 b_0^2W}{4a_0^2v} \biggl(\frac{2\pi a_0}{\beta v}\biggr)^{\frac{1}{2K}} \sin \biggl(\frac{\pi}{4K}\biggr) \biggl(\frac{\beta v}{2\pi}\biggr)^2
    \notag \\
    &\qquad \times B\biggl( \frac{1}{8K} - i\frac{\beta(\omega -v\theta_D)}{4\pi}, 1 - \frac{1}{4K}\biggr)
    \notag \\
    &\qquad \times B\biggl( \frac{1}{8K} - i\frac{\beta(\omega +v\theta_D)}{4\pi}, 1 - \frac{1}{4K}\biggr)
\end{align}
The DM interaction shifts $\omega$ in the retarded Green's function by $\pm v\theta_D$.
However, this shift does not change the imaginary part of $\Sigma_1^R(\gamma_g h_{\rm dc})$ (Appendix~\ref{app:dm}) when the temperature $T$ satisfies
\begin{align}
    \frac{1}{4\pi T} \max \{|\gamma_g h_{\rm dc}-v\theta_D|, |\gamma_g h_{\rm dc}+v\theta_D| \} \ll 1.
    \label{temp_range}
\end{align}
This condition \eqref{temp_range} on the temperature requires low magnetic fields and weak magnetic anisotropies.
Since CsCuCl$_3$ has weak magnetic anisotropies~\cite{Tanaka1992_CsCuCl3_ESR,Nihongi2022_CsCuCl3}, the condition \eqref{temp_range} will be experimentally feasible.

\subsection{Absence of ESR in  ferromagnetic spin chain}

The Gilbert damping $\alpha_{\rm G}$ is observable as the linewidth $\alpha_{\rm G} \gamma_g h_{\rm dc}$ of the ESR spectrum $I_{\rm 1d} (\omega)$ [Eq.~\eqref{ESR_spec_ours_def}] due to the dynamical correlation within the spin chain.
We reviewed in Sec.~\ref{sec:review_esr} that the easy-plane \emph{antiferromagnetic} spin chain yields the ESR peak with resonance frequency $\omega =\gamma'_g h_{\rm dc}$ and the linewidth $\propto T$~\cite{oshikawa2002_esr}.
This ESR could potentially mask the FMR if $\gamma'_g\approx \gamma_g$.
Fortunately, however, this masking in our ferromagnetic chain does not occur for the following reason.
When we rewrote the quantum ferromagnetic spin chain as the TL liquid, we mapped the ferromagnetic chain into the anisotropic antiferromagnetic chain by performing the staggered rotation \eqref{pi_rot_sz}.
As we reviewed, the ESR is attributed to the mode with zero wavenumber $q=0$.
Due to the staggered rotation, the $q=0$ mode of our ferromagnetic spin chain is equivalent to the $q=\pi/a_0$ mode of the antiferromagnetic spin chain.
The latter is unrelated to the above-mentioned ESR of the quantum antiferromagnetic spin chain.
In addition, we can explicitly calculate the retarded Green's function and confirm the absence of the resonance at the frequency near the FMR one $\omega \approx \gamma_g h_{\rm dc}$.

\section{Summary and discussions}\label{sec:summary}

We investigated spin pumping due to FMR into the easy-plane quantum ferromagnetic spin chain compound.
We considered the junction of the one-dimensional spin chain compound and the three-dimensional ferromagnetic insulator as schematically depicted in Fig.~\ref{fig:setup}.
The property of the junction is controlled by two parameters: the junction length $W$ and the interfacial randomness (see Sec.~\ref{sec:hamiltonian}).
The key idea of our study is the use of the easy-plane quantum ferromagnetic spin chain, which is effectively described by the TL liquid with the extremely large Luttinger parameter $K$ as discussed in Secs.~\ref{sec:TL_large-K} and \ref{sec:bosonization} [see Eq.~\eqref{K_large}].
In this study, we call this large-$K$ TL liquid the strongly interacting TL liquid because the Luttinger parameter represents the strength of the interaction after mapping it into a one-dimensional interacting fermion system.
This strongly interacting TL liquid, which is hardly realized in other quantum anti-ferromagnetic spin chains or one-dimensional electron systems, can be realized by introducing the weak easy-plane magnetic anisotropy to the ferromagnetic spin chain.

The information of the strongly interacting TL liquid is encoded in the Gilbert damping $\delta \alpha_{\rm G}$ of the FMR.
When the interfacial interaction $\mathcal H_{\rm int}$ is perturbative to the other part of the Hamiltonian \eqref{H_tot_def}, the Gilbert damping \eqref{alpha_self-energy} is governed by the dynamical correlation within the spin chain.
We analytically derived the dynamical correlation [the retarded Green's function $G^R_{s^+s^-}(\omega)$] by taking advantage of the conformal symmetry of the TL liquid.

The main result of this paper, the parameter dependence of the Gilbert damping, is summarized in Fig.~\ref{fig:alpha}.
When the interface is clean enough, the Gilbert damping $\delta \alpha_{\rm G}$ grows rapidly as the temperature decreases.
The growth follows the power law [Eq.~\eqref{alpha_G1_power}].
 Gilbert damping also shows the power law dependence on junction length $W$.
The power differs depending on whether the junction length is shorter or longer than the thermal length [Eq.\eqref{thermal_length}].

We discussed the possible experimental realization of our theoretical setup in Sec.~\ref{sec:experiment}.
CsCuCl$_3$ is a promising candidate suitable for the quantum ferromagnetic spin chain.
In fact, the Hamiltonian \eqref{H_CsCuCl3_1d} proposed for this compound is the same as ours except for the uniform DM interaction.
We showed that the uniform DM interaction affects the wavenumber of the retarded Green's function but is likely to be irrelevant to the Gilbert damping.

\section*{Acknowledgments}

This work is supported by the Japan Society for the Promotion of Science (JSPS KAKENHI Grants Nos. JP20K03769, JP21H04565, JP21K03465, JP21H01800, JP23H01839, JP24K06951, and JP24H00322).
This work was partially supported by JST CREST Grant No. JPMJCR19J4, Japan, by the National Natural Science Foundation of China (NSFC) under Grant No. 12374126, and by the Priority Program of Chinese Academy of Sciences under Grant No. XDB28000000. 

\appendix 

\section{Gilbert damping and retarded Green's function}\label{app:LLG_to_GR}

Here, we derive the phenomenological form \eqref{GR_+_LLG} and \eqref{GR_-_LLG} of the retarded Green's functions $G^R_{S^+S^-}(\omega)$ and $G^R_{S^-S^+}(\omega)$ from the LLG equation \eqref{LLG_def}, namely,
\begin{align}
    \frac{dS_{\bm 0}^+}{dt}
    &= -i\gamma_g h_{\rm dc} S_{\bm 0}^+ - i\frac{\alpha_{\rm G}}{S_0}\biggl(
    S_{\bm 0}^z \frac{dS_{\bm 0}^+}{dt} - S_{\bm 0}^+ \frac{dS_{\bm 0}^z}{dt}
    \biggr), \\
    \frac{dS_{\bm 0}^-}{dt}
    &= i\gamma_g h_{\rm dc} S_{\bm 0}^- - i\frac{\alpha_{\rm G}}{S_0}\biggl(
    -S_{\bm 0}^z \frac{dS_{\bm 0}^-}{dt} + S_{\bm 0}^- \frac{dS_{\bm 0}^z}{dt}\biggr), \\
    \frac{dS_{\bm 0}^z}{dt} 
    &=-i\frac{\alpha_{\rm G}}{2S_0} \biggl(S_{\bm 0}^+ \frac{dS_{\bm 0}^-}{dt} - S_{\bm 0}^- \frac{dS_{\bm 0}^+}{dt} \biggr).
\end{align}
We rewrite the LLG equation in terms of the creation and annihilation operators of magnons,
\begin{align}
    S_{\bm 0}^+ 
    &= \sqrt{2S_0}b_{\bm 0}, \\
    S_{\bm 0}^-
    &= \sqrt{2S_0}b_{\bm 0}^\dag, \\
    S_{\bm 0}^z &= S_0 - b_{\bm 0}^\dag b_{\bm 0}.
\end{align}
Plugging these relations to the LLG equation, we obtain
\begin{align}
    \frac{db_{\bm 0}}{dt}
    &=-i\gamma_g h_{\rm dc} b_{\bm 0} 
    \notag \\
    &\qquad 
    - i\alpha_{\rm G} \biggl(1-\frac{b_{\bm 0}^\dag b_{\bm 0}}{S_0}\biggr)\frac{db_{\bm 0}}{dt}--\frac{\alpha_{\rm G}}{S_0} b_{\bm 0} \frac{d(b_{\bm 0}^\dag b_{\bm 0})}{dt}, \\
    \frac{db_{\bm 0}^\dag}{dt}
    &= i\gamma_g h_{\rm dc} b_{\bm 0}^\dag
    \notag \\
    &\qquad +i\alpha_{\rm G} \biggl( 1-\frac{b_{\bm 0}^\dag b_{\bm 0}}{S_0} \biggr) \frac{db_{\bm 0}^\dag}{dt} -i\frac{\alpha_{\rm G}}{S_0} b_{\bm 0}^\dag \frac{d(b_{\bm 0}^\dag b_{\bm 0})}{dt}, \\
    \frac{d(b_{\bm 0}^\dag b_{\bm 0})}{dt}
    &= i\alpha_{\rm G} \biggl( b_{\bm 0} \frac{db_{\bm 0}^\dag}{dt} - b_{\bm 0}^\dag \frac{db_{\bm 0}}{dt}
    \biggr).
\end{align}
In the ferromagnetic phase at low temperatures, the magnon density $\braket{b_{\bm 0}^\dag b_{\bm 0}}$ satisfy
\begin{align}
    \frac{\braket{b_{\bm 0}^\dag b_{\bm 0}}}{2S_0} \ll 1.
\end{align}
Discarding the small terms leads to
\begin{align}
    \frac{db_{\bm 0}}{dt}
    &\approx \frac{-i\gamma_g h_{\rm dc}}{1+i\alpha_{\rm G}} b_{\bm 0}, \\
    \frac{db_{\bm 0}^\dag}{dt} 
    &\approx \frac{i\gamma_g h_{\rm dc}}{1-i\alpha_{\rm G}} b_{\bm 0}^\dag.
\end{align}
We are now ready to calculate the retarded Green's functions.
\begin{align}
    G^R_{S^+S^-}(\omega) &\approx -2S_0 i \int_0^\infty dt \, e^{i\omega t} \braket{[b_{\bm 0}(t), b_{\bm 0}^\dag(0)]}, \\
    G^R_{S^-S^+}(\omega)
    &\approx -2S_0i\int_0^\infty dt \, e^{i\omega t} \braket{[b_{\bm 0}^\dag(t), b_{\bm 0}(0)]}.
\end{align}
Integrating by parts, we find
\begin{align}
    G^R_{S^+S^-}(\omega)
    &= \frac{2S_0\braket{b_{\bm 0}(0)b_{\bm 0}^\dag(0)}}{\omega}
    \notag \\
    &\qquad -\frac{2S_0}{\omega}\int_0^\infty dt\, e^{i\omega t} \biggl\langle \biggl[
    \frac{db_{\bm 0}(t)}{dt}, b_{\bm 0}^\dag (0)
    \biggr]\biggr\rangle
    \notag \\
    &= \frac{2S_0\braket{b_{\bm 0}(0)b_{\bm 0}^\dag(0)}}{\omega}
    \notag \\
    &\qquad + \frac{\gamma_gh_{\rm dc}}{\omega(1+i\alpha_{\rm G})}G^R_{S^+S^-}(\omega).
\end{align}
We thus obtain
\begin{align}
    G^R_{S^+S^-}(\omega)
    &\approx \frac{2S_0\braket{b_{\bm 0}b_{\bm 0}^\dag}(1+i\alpha_{\rm G})}{\omega -\gamma_g h_{\rm dc}+i\alpha_{\rm G}\omega},
    \label{GR_S+S-_pre}
\end{align}
and similarly,
\begin{align}
    G^R_{S^-S^+}(\omega)
    &\approx \frac{2S_0\braket{b_{\bm 0}^\dag b_{\bm 0}}(1-i\alpha_{\rm G})}{\omega+\gamma_gh_{\rm dc}-i\alpha_{\rm G}\omega}.
    \label{GR_S-S+_pre}
\end{align}
The ESR spectrum has the resonant peak at $\omega=\gamma_g h_{\rm dc}$ that signals the FMR.
Note that the Gilbert damping coefficient of the numerators of Eqs.~\eqref{GR_S+S-_pre} and \eqref{GR_S-S+_pre} is negligible when the FMR frequency $\omega\approx\gamma_gh_{\rm dc}$ is concerned.

\section{Details of integration}

\subsection{Clean part in short-junction limit}\label{app:clean_short}

Here we describe technical details in the calculation of the self energy,
\begin{align}
    \Sigma_1^R(\omega)
    &= \frac{N_c\mathcal J_1^2b_0^2}{2a_0^2}\int_0^\infty dt \, e^{i\omega t} \int_0^W dx \int_0^W dy \im g(t,x-y),
    \label{Self-energy_1_app}
\end{align}
with 
\begin{align}
    g(t,z)
    &= \biggl(\frac{\sinh(i\pi a_0/\beta v)}{\sinh[\pi(ia_0-z-vt)/\beta v]}\biggr)^{\frac{1}{4K}}
    \notag \\
    &\qquad \times 
    \biggl(\frac{\sinh(i\pi a_0/\beta v)}{\sinh[\pi(ia_0+z-vt)/\beta v]}\biggr)^{\frac{1}{4K}}.
\end{align}
Note that $g(t,z)$ satisfies $g(t,z)^\dag = g(-t,z)$.
In other words,
\begin{align}
    -\im g(t,z) = \im g(-t,z).
\end{align}
Using this relation, we may rewrite the temporal integral as follows.
\begin{align}
    \Sigma_1^R(\omega)
    &= i\frac{N_c\mathcal J_1^2b_0^2}{2a_0^2}\int_0^\infty dt \, \sin(\omega t)
    \notag \\
    &\qquad \times \int_0^W dx \int_0^W dy \im g(t,x-y).
\end{align}

To simplify the notation, we first rescale the parameters.
\begin{align}
     \omega' &= \frac{\beta}{\pi} \omega, \quad \alpha = \frac{\pi}{\beta v}a_0, \quad w= \frac{\pi}{\beta v} W, \notag \\
    x' &= \frac{\pi}{\beta v} x, \quad y' = \frac{\pi}{\beta v}y \, \quad t' = \frac{\pi}{\beta}t.
\end{align}
This rescaling simplifies $g(t,z)$
\begin{align}
    g(t,z)
    = \biggl(\frac{\sinh(i\alpha)}{\sinh(i\alpha-z'-t')} 
    \biggr)^{\frac{1}{4K}}
    \biggl(
    \frac{\sinh(i\alpha)}{\sinh(i\alpha+z'-t')}
    \biggr)^{\frac{1}{4K}},
\end{align}
where $z'=\pi z/\beta v$.
The self energy then becomes
\begin{align}
    \Sigma_1^R(\omega)
    &= i\frac{N_c\mathcal J_1^2b_0^2 v^2}{2a_0^2} \biggl(\frac{\beta}{\pi}\biggr)^3\int_0^\infty dt' \sin(\omega' t')
    \notag \\
    &\qquad \times \int_0^w dx' \int_0^w dy' \im g(t,x-y)
    \label{Self-energy_1_rescaled}
\end{align}
Following Ref.~\cite{Fukuzawa2023}, we transform the spatial variables so that
\begin{align}
    Z = \frac{x'+y'}{2}, \quad z' =x'-y',
\end{align}
leading to
\begin{widetext}
\begin{align}
    \int_0^\infty dt' \sin(\omega't') \int_0^w dx' \int_0^w dy' \im g(t,x-y)
    &= - \frac 12 \int_{-\infty}^\infty dt' \int_0^{w/2}dZ \int_{-2Z}^{2Z} dz' (e^{i\omega' t'} - e^{-i\omega' t'}) g(t,x-y).
\end{align}
and
\begin{align}
    \Sigma_1^R(\omega)
    &= -\frac{N_c\mathcal J_1^2b_0^2 v^2}{4a_0^2}\biggl(\frac{\beta}{\pi}\biggr)^3
    \int_0^{w/2} dZ \int_{-2Z}^{2Z} dz' \Bigl[
    \mathcal C_{1/4K}(\omega',z') - \mathcal C_{1/4K}(-\omega',z')
    \Bigr],
    \label{Self-energy_1_C}
\end{align}
with
\begin{align}
    \mathcal C_{\gamma}(\zeta,z)
    = \int_{-\infty}^\infty du \, e^{i\zeta u} \biggl(\frac{\sinh(i\alpha)}{\sinh(i\alpha+z-u)}\biggr)^{\gamma}\biggl(\frac{\sinh(i\alpha)}{\sinh(i\alpha-z-u)}\biggr)^{\gamma}.
\end{align}
In this short-junction limit, we may approximate $\mathcal C_{\gamma}(\zeta,z) \approx \mathcal C_{\gamma}(\zeta,0)$ because $w=W/\ell_{\rm th} \ll 1$.
Then, we obtain
\begin{align}
    \mathcal C_{\gamma}(\zeta,0)
    &= \int_{-\infty}^\infty du \, e^{i\zeta u} \biggl(\frac{\sinh(i\alpha)}{\sinh(i\alpha-u)}\biggr)^{2\gamma}
    \notag \\
    &= (1-e^{-2i\alpha})^{2\gamma} \int_{-\infty}^\infty du \, e^{-u(2\gamma-i\zeta)} \frac{1}{[e^{-2u}+e^{i(\pi-2\alpha)}]^{2\gamma}} .
\end{align}
Here, we apply the following integration formula (3.314 in Ref.~\cite{zwillinger2007table_integrals})
\begin{align}
    \int_{-\infty}^\infty dx \, \frac{e^{-\mu x}}{(e^{-x/c}+ e^{b/c})^\nu} = c \exp\biggl[
    b \biggl(\mu - \frac{\nu}{c}\biggr)
    \biggr]B (c \mu, \, \nu-c \mu),
\end{align}
valid for 
\begin{align}
    \re \biggl( \frac{\nu}{c} \biggr) > \re \mu > 0, \quad |\im b| < \pi \re c.
    \label{cond_integration_formula}
\end{align}
We choose
\begin{align}
    b =  \frac{i}{2}(\pi-2\alpha), \quad
    c = \frac 12, \quad
    \mu =2 \gamma - i\zeta, \quad
    \nu = 2\gamma.
\end{align}
This choice of parameters meets condition \eqref{cond_integration_formula} for $\zeta>0$, $\alpha>0$, and $\gamma>0$.
We thus obtain the following.
\begin{align}
    \mathcal C_{\gamma}(\zeta,0)
    &= (1-e^{-2i\alpha})^{2\gamma} \frac 12 \exp\biggl[
    \frac i2(\pi-2\alpha) (-2\gamma-i\zeta)
    \biggr]B\biggl(\gamma -i\frac{\zeta}{2}, \, \gamma + i\frac{\zeta}{2}
    \biggr)
    \notag \\
    &= \frac 12 (1-e^{-2i \alpha})^{2\gamma}e^{-i\pi\gamma} e^{\frac{\pi\zeta}{2}}\frac{|\Gamma(\gamma +i\frac{\zeta}{2})|^2}{\Gamma(2\gamma)}
    \notag \\
    &\approx \frac 12 (2 \alpha)^{2\gamma}  e^{\frac{\pi\zeta}{2}}\frac{|\Gamma(\gamma +i\frac{\zeta}{2})|^2}{\Gamma(2\gamma)}.
\end{align}
In the last line, we used $\alpha \ll 1$ valid in the TL-liquid phase.

Plugging this result in Eq.~\eqref{Self-energy_1_C}, we obtain
\begin{align}
    \Sigma_1^R(\omega)
    &= - \frac{N_c\mathcal J_1^2b_0^2v^2}{4a_0^2} \biggl(\frac{\beta}{\pi}\biggr)^3\frac{w^2}2 \frac 12(2\alpha)^{\frac{1}{2K}}(e^{i\frac{\pi\omega'}{2}} - e^{-i\frac{\pi\omega'}{2}})\frac{|\Gamma(\gamma +i\frac{\zeta}{2})|^2}{\Gamma(2\gamma)}
    \notag \\
    &\approx -i\frac{N_c\mathcal J_1^2b_0^2v^2}{16a_0^2}  \biggl(\frac{\beta}{\pi}\biggr)^3 \biggl(\frac{\pi W}{\beta v}\biggr)^2 \biggl(\frac{2\pi a_0}{\beta v}\biggr)^{\frac{1}{2K}}  \beta \omega \frac{\Gamma(\frac{1}{4K})^2}{\Gamma(\frac{1}{2K})} \notag \\
    &\approx -i\frac{\pi N_c\mathcal J_1^2b_0^2W^2}{4v^2} \omega \biggl(\frac{2\pi a_0}{\beta v}\biggr)^{\frac 1{2K}-2}\frac{\Gamma(\frac{1}{4K})^2}{\Gamma(\frac{1}{2K})},
\end{align}
where we assumed $\beta \omega =\omega/T \ll 1$.
This assumption is justified for weak magnetic fields $\gamma_g h_{\rm dc}/T \ll 1$ when the FMR with resonance frequency $\omega=\gamma_g h_{\rm dc}$ is concerned.
Therefore, we arrive at the following representation of the imaginary part.
\begin{align}
    \im \Sigma_1^R(\omega) &\approx -\frac{\pi N_c\mathcal J_1^2b_0^2W^2}{4v^2} \omega \biggl(\frac{2\pi a_0}{\beta v}\biggr)^{\frac 1{2K}-2}\frac{\Gamma(\frac{1}{4K})^2}{\Gamma(\frac{1}{2K})}
    \label{app_self-energy_clean_short}
\end{align}
When $K\gg 1$, this expression is further simplified as
\begin{align}
    \im \Sigma_1^R(\omega)
    &\approx -\frac{2\pi N_c\mathcal J_1^2b_0^2W^2K}{v^2} \omega \biggl(\frac{2\pi a_0}{\beta v}\biggr)^{\frac 1{2K}-2}.
    \label{app_self-energy_clean_short_large_K}
\end{align}

\subsection{Clean part in long-junction limit}

We derive the same self-energy $\Sigma_1^R(\omega)$ in the long-junction limit $W \gg \ell_{\rm th}$.
The integrand $\im g(t,z)$ of the self-energy \eqref{Self-energy_1_app} is bound within the light cone $|z|< vt$~\cite{Giamarchi2003}:
\begin{align}
    \im g(t,z)
    &= -\theta_{\rm H}(vt-x)\theta_{\rm H}(vt+x)\biggl(\frac{\pi a_0}{\beta v}\biggr)^{\frac{1}{2K}} \sin\biggl(\frac{\pi}{4K}\biggr)\biggl(\frac{1}{\sinh[\pi(vt+z)/\beta v]}\biggr)^{\frac{1}{4K}} \biggl(\frac{1}{\sinh[\pi(vt-z)/\beta v]}\biggr)^{\frac{1}{4K}}.
\end{align}
Since the right-hand side exponentially decays and almost vanishes for $|z|\gg \ell_{\rm th}$, we can approximate the self-energy \eqref{Self-energy_1_app} as
\begin{align}
    \Sigma_1^R(\omega)
    &\approx -\frac{N_c\mathcal J_1^2b_0^2W}{2a_0^2}\int_0^\infty dt \, e^{i\omega t}  \int_{0}^\infty dz \, \im g(t,z).
\end{align}
Transforming the spatial variables $\xi_\pm = vt\pm z$,  we can simplify the integrals as
\begin{align}
    \Sigma_1^R(\omega)
    &= -\frac{N_c\mathcal J_1^2b_0^2W}{4a_0^2v}\biggl(\frac{\pi a_0}{\beta v}\biggr)^{\frac{1}{2K}} \sin \biggl(\frac{\pi}{4K}\biggr) \int_0^\infty d\xi_+ e^{i\omega\xi_+/2v} \biggl(\frac{1}{\sinh(\pi \xi_+/\beta v)}\biggr)^{\frac{1}{4K}} \int_0^\infty d\xi_-e^{i\omega \xi_-/2v} \biggl(\frac{1}{\sinh(\pi \xi_-/\beta v)}\biggr)^{\frac{1}{4K}}
    \notag \\
    &= - \frac{N_c\mathcal J_1^2b_0^2W}{4a_0^2v} \biggl(\frac{2\pi a_0}{\beta v}\biggr)^{\frac{1}{2K}} \sin\biggl(\frac{\pi}{4K}\biggr) \biggl(\frac{\beta v}{2\pi}\biggr)^2 \biggl[
    B\biggl(\frac{1}{8K}-i\frac{\beta\omega}{4\pi}, 1-\frac{1}{4K}\biggr)
    \biggr]^2,
\end{align}
where we used the integration formula~\cite{Giamarchi2003},
\begin{align}
    \int_0^\infty d\xi e^{i\omega \xi/2v} \biggl(\frac{1}{\sinh(\pi \xi/\beta v)}\biggr)^{\frac{1}{4K}} &= 2^{\frac{1}{4K}} \frac{\beta v}{2\pi } B\biggl(\frac{1}{8K} -i\frac{\beta v}{4\pi}, 1-\frac{1}{4K}\biggr).
\end{align}
Due to the identity $\Gamma(z)\Gamma(1-z)=\pi/\sin(\pi z)$ for the complex $z$,
the Beta function on the right-hand side reads as
\begin{align}
    B\biggl(\frac{1}{8K}-i\frac{\beta\omega}{4\pi}, 1-\frac{1}{4K}\biggr) = \frac{1}{\sin(\frac{\pi}{4K})\Gamma(\frac{1}{4K})} \biggl|\Gamma\biggl(\frac{1}{8K}-i\frac{\beta\omega}{4\pi}\biggr)\biggr|^2\sin \biggl(\frac{\pi}{8K} + i\frac{\beta\omega}{4}\biggr).
\end{align}
We thus obtain the clean part in the long-junction limit.
\begin{align}
    \Sigma_1^R(\omega)
    &= - \frac{N_c\mathcal J_1^2b_0^2W}{4v} \biggl(\frac{2\pi a_0}{\beta v}\biggr)^{\frac 1{2K}-2} \frac{1}{\sin^2(\frac{\pi}{4K})\Gamma(\frac{1}{4K})^2} \biggl|\Gamma\biggl(\frac{1}{8K}-i\frac{\beta\omega}{4\pi} \biggr)\biggr|^4 \sin^2 \biggl(\frac{\pi}{8K}+i\frac{\beta\omega}{4}\biggr).
\end{align}
For later convenience, we take its imaginary part for $\beta\omega \ll 1$.
\begin{align}
    \im \Sigma_1^R(\omega) &\approx  -\frac{N_c\mathcal J_1^2b_0^2 W}{16v} \beta\omega \biggl(\frac{2\pi a_0}{\beta v}\biggr)^{\frac{1}{2K}-2} \biggl(\frac{\Gamma(\frac{1}{8K})^2}{\Gamma(\frac{1}{4K})}\biggr)^2 .
    \label{app_self-energy_clean_long}
\end{align}
For $K\gg 1$, 
\begin{align}
    \im \Sigma_1^R(\omega) &\approx - \frac{N_c\mathcal J_1^2b_0^2WK^2}{\pi v} \beta \omega \biggl(\frac{2\pi a_0}{\beta v}\biggr)^{\frac{1}{2K}-2}.
    \label{app_self-energy_clean_long_large_K}
\end{align}

\subsection{Dirty part}\label{app:dirty}

The dirty part $\Sigma_2^R(\omega)$ is simpler than the clean one, since the former does not involve spatial integration.
\begin{align}
    \Sigma_2^R(\omega)
    &= -i\frac{N_c\mathcal J_2b_0^2 W}{4a_0}\int_0^\infty dt \, e^{i\omega t} \braket{[e^{-i\theta(t,0)}, e^{i\theta(0,0)}]}
    \notag \\
    &= \frac{N_c\mathcal J_2b_0^2 W}{2a_0} \int_0^\infty dt \, e^{i\omega t} \im g(t,0).
\end{align}
The same temporal integration has already been done in Appendix~\ref{app:clean_short}.
\begin{align}
    \int_0^\infty dt \, e^{i\omega t} \im g(t,0)
    &=  - \frac 14 \int_{-\infty}^\infty dt \, (e^{i\omega t} - e^{-i\omega t}) g(t,0)
    \notag \\
    &= - \frac 14 \frac{\beta}{\pi} \Bigl[
    \mathcal C_{1/4K}(\omega',0) - \mathcal C_{1/4K}(-\omega',0)
    \Bigr]
    \notag \\
    &= - \frac 18 \frac{\beta}{\pi} \biggl(\frac{2\pi a_0}{\beta v}\biggr)^{\frac{1}{2K}} (e^{i\beta \omega}-e^{-i\beta \omega}) \frac{|\Gamma(\frac{1}{4K}+i\frac{\beta\omega}{2\pi})|^2}{\Gamma(\frac{1}{2K})}
    \notag \\
    &\approx - i\frac{\pi a_0^2}{v^2} \omega \biggl(\frac{2\pi a_0}{\beta v}\biggr)^{\frac{1}{2K}-2} \frac{\Gamma(\frac{1}{4K})^2}{\Gamma(\frac 1{2K})},
\end{align}
where we used $\beta\omega\ll 1$ in the last line.
We thus obtain
\begin{align}
    \Sigma_2^R(\omega) &\approx - i\frac{\pi N_c \mathcal J_2b_0^2a_0 W}{2v^2} \omega \biggl(\frac{2\pi a_0}{\beta v}\biggr)^{\frac 1{2K}-2} \frac{\Gamma(\frac{1}{4K})^2}{\Gamma(\frac{1}{2K})}.
        \label{app_self-energy_dirty}
\end{align}
In the large-$K$ limit, we can approximate it as
\begin{align}
    \Sigma_2^R(\omega) &\approx -i\frac{4\pi N_c\mathcal J_2b_0^2 a_0WK}{v^2} \omega \biggl(\frac{2\pi a_0}{\beta v}\biggr)^{\frac{1}{2K}-2}
        \label{app_self-energy_dirty_large_K}
\end{align}

\subsection{DM interaction}\label{app:dm}

As we discussed in the main text, the uniform DM interaction affects only the clean part in the long-junction limit.
The self-energy $\Sigma_1^R(\omega)$ is then given by
\begin{align}
    \Sigma_1^R(\omega)
    &= - \frac{N_c \mathcal J_1^2 b_0^2W}{4a_0^2v} \biggl(\frac{2\pi a_0}{\beta v}\biggr)^{\frac{1}{2K}} \sin \biggl(\frac{\pi}{4K}\biggr) \biggl(\frac{\beta v}{2\pi}\biggr)^2
    \notag \\
    &\qquad \times B\biggl( \frac{1}{8K} - i\frac{\beta(\omega -v\theta_D)}{4\pi}, 1 - \frac{1}{4K}\biggr)
    B\biggl( \frac{1}{8K} - i\frac{\beta(\omega +v\theta_D)}{4\pi}, 1 - \frac{1}{4K}\biggr).
    \label{Sigma1_DM}
\end{align}
Here, we show that the imaginary part of Eq.~\eqref{Sigma1_DM} is independent of $\theta_D = \tan^{-1}(D/2J_0)$, when the temperature $T=1/\beta$ satisfies
\begin{align}
    \frac{\beta}{4\pi} \max \{|\omega-v\theta_D|, |\omega+v\theta_D| \} \ll 1.
    \label{cond_DM_indep}
\end{align}
Let us rewrite the product of the beta functions as follows.
\begin{align}
    &B\biggl( \frac{1}{8K} - i\frac{\beta(\omega -v\theta_D)}{4\pi}, 1 - \frac{1}{4K}\biggr)
    B\biggl( \frac{1}{8K} - i\frac{\beta(\omega +v\theta_D)}{4\pi}, 1 - \frac{1}{4K}\biggr)
    \notag \\
    &= \biggl|\Gamma \biggl(\frac{1}{8K} -i\frac{\beta(\omega-v\theta_D)}{4\pi}\biggr)\Gamma \biggl(\frac{1}{8K} -i\frac{\beta(\omega+v\theta_D)}{4\pi}\biggr)\biggr|^2 \frac{1}{\sin^2(\frac{\pi}{4K}) \Gamma(\frac{1}{4K})^2}
    \notag \\
    &\qquad \times \sin\biggl(\frac{\pi}{8K} +i\frac{\beta(\omega-v\theta_D)}{4}\biggr)\sin\biggl(\frac{\pi}{8K} +i\frac{\beta(\omega+v\theta_D)}{4}\biggr)
    \notag \\
    &= \biggl|\Gamma \biggl(\frac{1}{8K} -i\frac{\beta(\omega-v\theta_D)}{4\pi}\biggr)\Gamma \biggl(\frac{1}{8K} -i\frac{\beta(\omega+v\theta_D)}{4\pi}\biggr)\biggr|^2 \frac{1}{\sin^2(\frac{\pi}{4K}) \Gamma(\frac{1}{4K})^2}
    \notag \\
    &\qquad \times \frac 12 \biggl[
    \cosh\biggl(\frac{\beta v\theta_D}{2\pi}\biggr) - \cos \biggl( \frac{\pi}{4K} + i\frac{\beta \omega}{2}\biggr)
    \biggr].
\end{align}
The imaginary part $\im \Sigma_1^R(\omega)$ thus becomes
\begin{align}
   \im \Sigma_1^R(\omega)
   &= - \frac{N_c \mathcal J_1^2 b_0^2W}{4a_0^2v} \biggl(\frac{2\pi a_0}{\beta v}\biggr)^{\frac{1}{2K}} \biggl(\frac{\beta v}{2\pi}\biggr)^2
   \notag \\
   &\qquad \times \biggl|\Gamma \biggl(\frac{1}{8K} -i\frac{\beta(\omega-v\theta_D)}{4\pi}\biggr)\Gamma \biggl(\frac{1}{8K} -i\frac{\beta(\omega+v\theta_D)}{4\pi}\biggr)\biggr|^2 \frac{1}{\Gamma(\frac{1}{4K})^2} \frac 12 \sinh \biggl(\frac{\beta\omega}{2}\biggr).
\end{align}
Under the condition \eqref{cond_DM_indep}, we can approximate it as
\begin{align}
    \im \Sigma_1^R(\omega)
    &\approx - \frac{N_c\mathcal J_1^2 b_0^2 W}{16v} \biggl(\frac{2\pi a_0}{\beta v}\biggr)^{\frac 1{2K}-2} \beta \omega \biggl(\frac{\Gamma(\frac{1}{8K})^2}{\Gamma(\frac{1}{4K})}\biggr)^2,
\end{align}
which is identical to Eq.~\eqref{app_self-energy_clean_long} evaluated in the absence of the DM interaction.

\end{widetext}

\end{document}